\documentclass[conference]{IEEEtran}

\usepackage{cite}
\usepackage{amsmath,amssymb,amsfonts}
\usepackage{booktabs}
\usepackage{multirow}
\usepackage{graphicx}
\usepackage[caption=false,font=footnotesize]{subfig}
\usepackage{siunitx}
\usepackage{textcomp}
\usepackage{xcolor}
\usepackage{xspace}
\usepackage{tikz}
\usepackage{algpseudocode}
\usetikzlibrary{arrows.meta}

\newcounter{algorithm}

\graphicspath{{./}}
\newcommand{\campa}{CAMPA\xspace}
\newcommand{\mlkem}{ML-KEM\xspace}
\newcommand{\mldsa}{ML-DSA\xspace}
\newcommand{\slhdsa}{SLH-DSA\xspace}
\newcommand{\riscv}{RISC-V\xspace}
\newcommand{\keccak}{Keccak-$f[1600]$\xspace}

\title{CAMPA: A Cross-Algorithm Multithreaded PQC Accelerator with Dynamic Scheduling}

\author{%
  \IEEEauthorblockN{Zhengpeng Zhao\IEEEauthorrefmark{2}, Wenbo Xu\IEEEauthorrefmark{2}, Jiayi Li, Dongsheng Liu\IEEEauthorrefmark{1}, Jiahao Lu\IEEEauthorrefmark{1}, Ang Hu, Huanyu Wang}
  \IEEEauthorblockA{Huazhong University of Science and Technology, Wuhan, China\\
    \IEEEauthorrefmark{1}Corresponding authors. \IEEEauthorrefmark{2}Joint first authors.}
}

\begin{document}

\maketitle

\begin{abstract}
  Multi-scheme post-quantum cryptography (PQC) accelerators share operators to
reduce hardware cost, but data dependencies and changing resource demands
within each task leave shared hardware idle. We present \campa, a \riscv PQC
coprocessor supporting \mlkem, \mldsa, and \slhdsa-SHAKE. To the best of our
knowledge, \campa is the first PQC coprocessor for \riscv CPUs to support hardware
dynamic scheduling across independent cryptographic requests. The hardware scheduler pools ready operations across independent tasks
according to runtime data and resource availability, filling idle execution windows.
Coordinated RAM version management and context
retention protect data and state during this interleaving, enabling
computation and data movement to overlap across tasks on shared datapaths
and storage. Architectural simulation of 128-request batches across 18
\mlkem/\mldsa workloads shows that the Main and Performance configurations
achieve geometric-mean throughput ratios of
$1.21\times$ and $1.59\times$, respectively, relative to serial execution of
the same task sets. In Main, mean polynomial-unit execution activity is
77.5\% for single requests and 93.3\% for these batches. Synthesized in SMIC
28~nm under an 800~MHz clock constraint, the Main configuration occupies
0.390~mm$^2$, including on-chip memory.
\end{abstract}


\section{Introduction}\label{sec:introduction}

Sharing hardware across multiple post-quantum cryptography (PQC) schemes offers an area-efficient basis for cryptographic acceleration. NIST's \mlkem, \mldsa, and \slhdsa standardize key encapsulation and signatures~\cite{nist-fips-203,nist-fips-204,nist-fips-205}. The lattice schemes share polynomial arithmetic, hashing, and sampling~\cite{kyber-original,dilithium-original}; \slhdsa-SHAKE shares their Keccak computation~\cite{riscv-keccak} and is denoted \slhdsa hereafter. These common operations enable hardware reuse across schemes~\cite{kali,unified-pqc-soc,unified-lattice-cryptoprocessor,scalable-multi-pqc}. Within a request, however, data dependencies and changing operator demands create idle execution windows.

Exploiting these windows requires both independent work and a mechanism to select it. Programmable processors and instruction extensions make accelerated primitives available to software~\cite{sapphire,vpqc,risqv,polarfire-codesign,opentitan-pqc,pqcuark}; hardware scoreboarding and dispatch to configurable arrays coordinate independent computation~\cite{matrix-riscv,agile-pqc}, while unified cryptoprocessors overlap polynomial and hash execution~\cite{kali,unified-pqc-soc,unified-lattice-cryptoprocessor}. FPGA batching and GPU signature scheduling further exploit independence across requests~\cite{high-volume-pqc,gpu-dilithium}. The opportunity is to combine this larger pool of independent work with operation-level hardware scheduling in a shared coprocessor.

\campa implements this approach in a \riscv PQC coprocessor. Each independent cryptographic request forms a \emph{task}, with its own execution state retained in a hardware thread. The scheduler selects ready operations across tasks according to operand dependencies and shared-resource availability. When one task waits for data or resources, ready work from another can fill idle execution windows. This interleaving supports concurrent progress across algorithms on shared compute and storage resources.

\begin{figure}[t]
  \centering
  \includegraphics[width=\columnwidth]{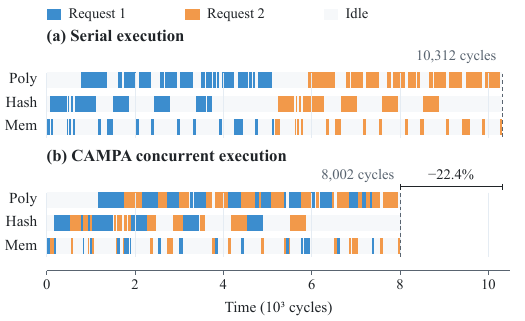}
  \caption{Two \mlkem-768 encapsulations on Main T2.}
  \label{fig:introduction-timeline}
\end{figure}

Interleaving two independent \mlkem-768 encapsulations reduces their total completion time by 22.4\% on the same two-thread Main configuration (Main T2; Fig.~\ref{fig:introduction-timeline}). Multiple requests supply ready work without replicating the accelerator~\cite{simultaneous-multithreading}.

Interleaving must preserve operand identities and protect data while arbitrating RAM ports and hash contexts. \textbf{To the best of our knowledge, CAMPA is the first PQC coprocessor for RISC-V CPUs to support hardware dynamic scheduling across independent cryptographic requests.} Its contributions are:

\begin{itemize}

\item \textbf{Cross-algorithm multithreaded dynamic scheduling.} Hardware threads keep independent \mlkem and \mldsa tasks in flight. Per-unit reservation stations and resource arbitration select issuable operations according to runtime operand readiness and resource availability, overlapping computation and data movement across tasks to improve resource utilization and task throughput.

\item \textbf{RAM versioning and dependency tracking.} We adapt Clockhands relative naming~\cite{clockhands} to RAM blocks. Operand binding and per-thread reader tracking preserve data identity and protect live versions during reordered execution and in-place updates. Physical-bank coordination manages concurrent tasks' access to shared storage.

\item \textbf{Operator execution over shared resources.} The number-theoretic transform (NTT) and its inverse (INTT) reuse slots across stages and regroup accesses to supply butterfly pairs from shared single-port RAM. Hash reuses Keccak state banks for independent lattice-hash contexts and \slhdsa work/park execution, enabling independent-call overlap and direct chain-endpoint consumption in incremental compression, reducing intermediate data movement and software coordination.

\end{itemize}

Across 18 \mlkem/\mldsa workloads, the Main and Performance configurations achieve $1.21\times$ and $1.59\times$ geometric-mean throughput over serial execution for 128-request batches. Main's mean polynomial-unit activity is 77.5\% for single requests and 93.3\% for batches. Its synthesized 28-nm area is 0.390~mm$^2$ including SRAM.
\section{Preliminaries}\label{sec:background}

\mlkem and \mldsa use 256-coefficient polynomials modulo 3329 and 8380417, respectively~\cite{nist-fips-203,nist-fips-204}. Coefficient-wise operations and independent polynomial entries expose parallelism. Polynomial multiplication is a major lattice-PQC bottleneck, motivating number-theoretic transform (NTT) and inverse NTT (INTT) acceleration~\cite{cfntt,compact-kyber,compact-dilithium}. The forward NTT has seven radix-2 butterfly stages in \mlkem and eight in \mldsa (Algorithm~\ref{alg:forward-ntt}). Butterflies within a stage are independent; successive stages depend on prior results and change coefficient pairings. INTT reverses the spans with inverse twiddles and normalization.

\begin{figure}[t]
  \refstepcounter{algorithm}
  \label{alg:forward-ntt}
  \hrule
  \smallskip
  \noindent\textbf{Algorithm \thealgorithm}\quad In-place forward NTT
  \smallskip
  \hrule
  \smallskip
  {\footnotesize
  \begin{algorithmic}[1]
    \Require Coefficients $a[0\ldots255]$, modulus $q$, twiddle table $Z$
    \Require $\ell_{\min}=2$ for \mlkem; $\ell_{\min}=1$ for \mldsa
    \Ensure $a$ contains the scheme's NTT representation
    \State $k\gets1$
    \For{$\ell=128,64,\ldots,\ell_{\min}$} \Comment{Stage}
      \For{$b=0,2\ell,\ldots,256-2\ell$} \Comment{Block}
        \State $\zeta\gets Z[k]$; $k\gets k+1$
        \For{$j=b,b+1,\ldots,b+\ell-1$} \Comment{Butterfly}
          \State $u\gets a[j]$
          \State $t\gets\zeta\cdot a[j+\ell]\bmod q$
          \State $a[j]\gets(u+t)\bmod q$
          \State $a[j+\ell]\gets(u-t)\bmod q$
        \EndFor
      \EndFor
    \EndFor
    \State \Return $a$
  \end{algorithmic}}
  \smallskip
  \hrule
\end{figure}

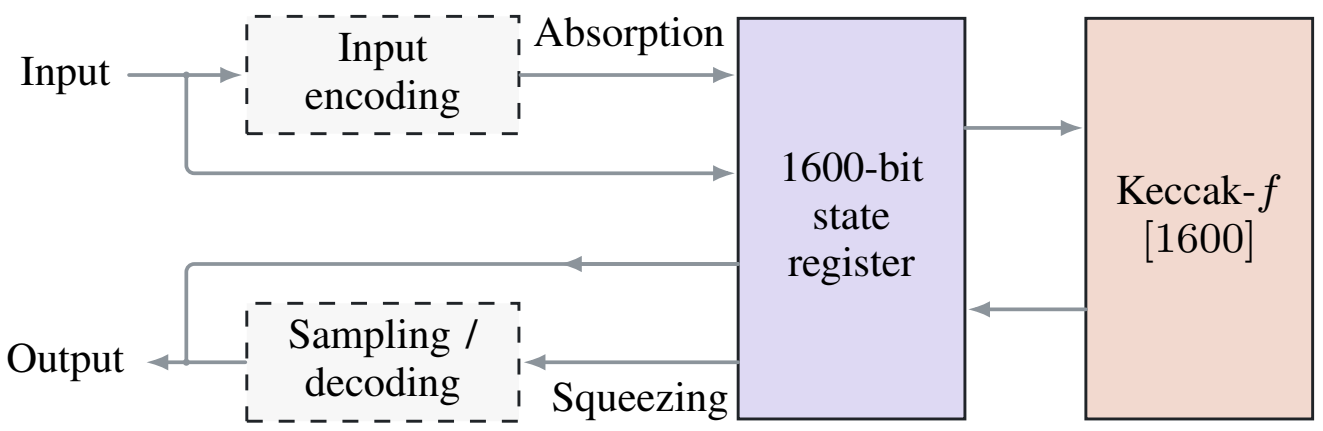
\begin{figure}[t]
  \centering
  \begin{tikzpicture}[
      font=\footnotesize,
      process/.style={draw=hashink, line width=0.55pt, align=center,
        minimum height=0.7cm, inner sep=3pt},
      optional/.style={process, dashed, fill=black!3},
      wire/.style={draw=hasharrow, line width=0.7pt,
        line cap=round, line join=round, rounded corners=1.5pt},
      flow/.style={wire, -{Latex[length=1.5mm,width=1.1mm]},
        shorten >=0.45pt}]
    \definecolor{hashink}{RGB}{32,36,40}
    \definecolor{hasharrow}{RGB}{140,148,155}
    \definecolor{hashstate}{RGB}{222,217,243}
    \definecolor{hashcompute}{RGB}{242,217,207}

    \node (input) at (0.1,0.95) {Input};
    \node[optional, minimum width=1.8cm] (encoding) at (2.2,0.95)
      {Input\\encoding};
    \node[process, fill=hashstate, minimum width=1.5cm,
      minimum height=2.65cm] (state) at (5.3,0)
      {1600-bit\\state\\register};
    \node[process, fill=hashcompute, minimum width=1.5cm,
      minimum height=2.65cm] (permute) at (7.6,0)
      {Keccak-$f$\\$[1600]$};
    \node[optional, minimum width=1.8cm] (sampling) at (2.2,-0.95)
      {Sampling /\\decoding};
    \node (output) at (0.1,-0.95) {Output};

    \draw[flow] (input) -- (encoding);
    \draw[flow] (encoding.east) --
      node[pos=0.5, above] {Absorption} (4.55,0.95);
    \draw[flow] (4.55,-0.95) --
      node[pos=0.45, below] {Squeezing} (sampling.east);
    \draw[wire] (sampling.west) -- (0.9,-0.95);
    \draw[flow] (0.9,-0.95) -- (output.east);

    \draw[flow] (0.9,0.95) -- (0.9,0.3) -- (4.55,0.3);
    \draw[wire] (4.55,-0.3) -- (0.9,-0.3) -- (0.9,-0.95);
    \draw[flow] (4.55,-0.3) -- (3.35,-0.3);
    \foreach \x/\y in {0.9/0.95,0.9/-0.95}
      \fill[hasharrow] (\x,\y) circle[radius=0.65pt];

    \draw[flow] (6.05,0.6) -- (6.85,0.6);
    \draw[flow] (6.85,-0.6) -- (6.05,-0.6);
  \end{tikzpicture}
  \caption{PQC hash dataflow~\cite{nist-fips-202}.}
  \label{fig:hash-dataflow}
\end{figure}

SHA3/SHAKE absorb blocks into a 1600-bit state and apply the 24-round \keccak permutation~\cite{nist-fips-202,kyber-sha3}; squeezing follows final absorption. Fig.~\ref{fig:hash-dataflow} shows optional encoding and sampling/decoding around these state updates, with bypasses for byte streams. Rejection sampling draws output until enough coefficients are accepted. Separate states let one call process input during another's permutation.

\slhdsa repeatedly evaluates short hashes to construct WOTS+ chains and FORS/tree nodes~\cite{nist-fips-205}. Node hashes reuse seeds and include domain-separation addresses (ADRS); WOTS+ chain endpoints and FORS roots then feed ordered multi-node compression. Compression $T_\ell$ can absorb nodes incrementally as they become available, retaining its SHAKE state while subsequent nodes are generated.

\begin{figure}[t]
  \centering
  \begin{tikzpicture}[
      >=Stealth, font=\fontsize{7.5}{8.6}\selectfont,
      operation/.style={draw=black!65,align=center,minimum height=0.65cm,
        inner sep=1.5pt},
      sample/.style={operation,fill=green!8,text width=1.05cm},
      compute/.style={operation,fill=violet!12,text width=1.26cm},
      check/.style={operation,fill=red!8,text width=0.82cm},
      flow/.style={-{Stealth[length=1.2mm,width=0.85mm]},
        draw=black!60,line width=0.55pt},
      lane/.style={draw=black!18,line width=0.4pt}]
    \draw[lane] (1.0,0) -- (8.25,0);
    \draw[lane] (1.0,-1.35) -- (8.25,-1.35);
    \node[align=center] at (0.43,0) {Request\\A};
    \node[align=center] at (0.43,-1.35) {Request\\B};

    \node[sample] (a0) at (1.55,0) {Sampling};
    \node[compute] (a1) at (2.95,0) {Polynomial};
    \node[check] (a2) at (4.35,0) {Checks};
    \node[sample] (a3) at (5.75,0) {Sampling};
    \node[compute] (a4) at (7.15,0) {Polynomial};
    \draw[flow] (a0.east) -- (a1.west);
    \draw[flow] (a1.east) -- (a2.west);
    \draw[flow,dashed] (a2.north) -- (4.35,0.65) --
      node[above,font=\scriptsize] {Fail: retry} (5.75,0.65) -- (a3.north);
    \draw[flow] (a3.east) -- (a4.west);
    \draw[flow] (a4.east) -- (8.0,0);
    \node at (8.17,0) {$\cdots$};

    \node[sample] (b0) at (2.95,-1.35) {Sampling};
    \node[compute] (b1) at (4.35,-1.35) {Polynomial};
    \node[check] (b2) at (5.75,-1.35) {Checks};
    \node[operation,fill=black!3,text width=1.05cm]
      (b3) at (7.25,-1.35) {Encode\\signature};
    \draw[flow] (b0.east) -- (b1.west);
    \draw[flow] (b1.east) -- (b2.west);
    \draw[flow] (b2.east) -- node[above=6pt,font=\scriptsize] {Pass} (b3.west);
  \end{tikzpicture}
  \caption{Schematic progression of independent \mldsa signing requests~\cite{nist-fips-204}.}
  \label{fig:request-dependencies}
\end{figure}
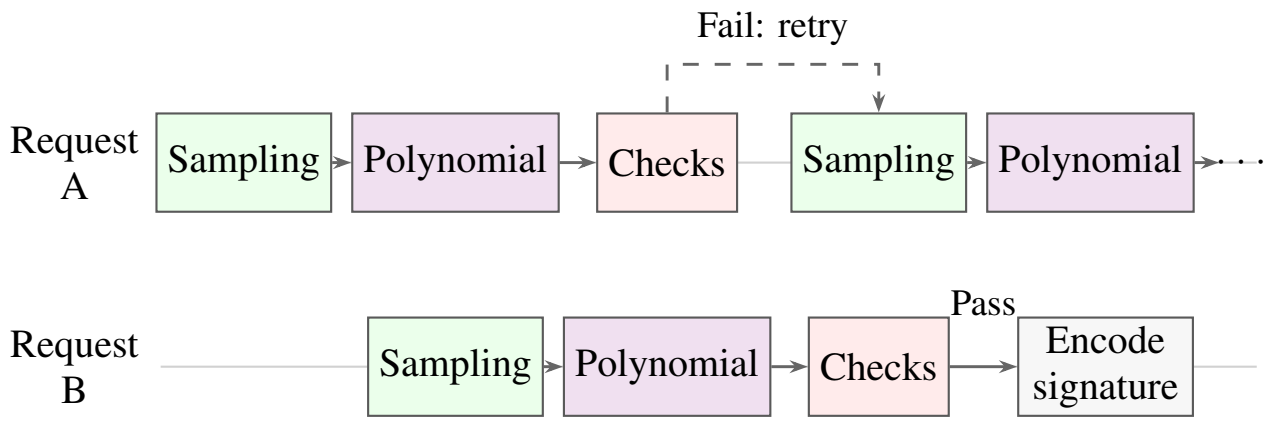

A request is one KeyGen, Encaps, Decaps, Sign, or Verify operation. Data dependencies and control decisions constrain its progress: \mldsa signing synchronizes norm and hint-count checks before accepting or retrying~\cite{nist-fips-204}. Independent requests can progress separately. Fig.~\ref{fig:request-dependencies} shows one signing request retrying while another finishes; sampling and polynomial work from different requests overlap on separate units.

\section{CAMPA Architecture}\label{sec:architecture}

\begin{figure*}[t]
  \centering
  \includegraphics[width=0.96\textwidth]{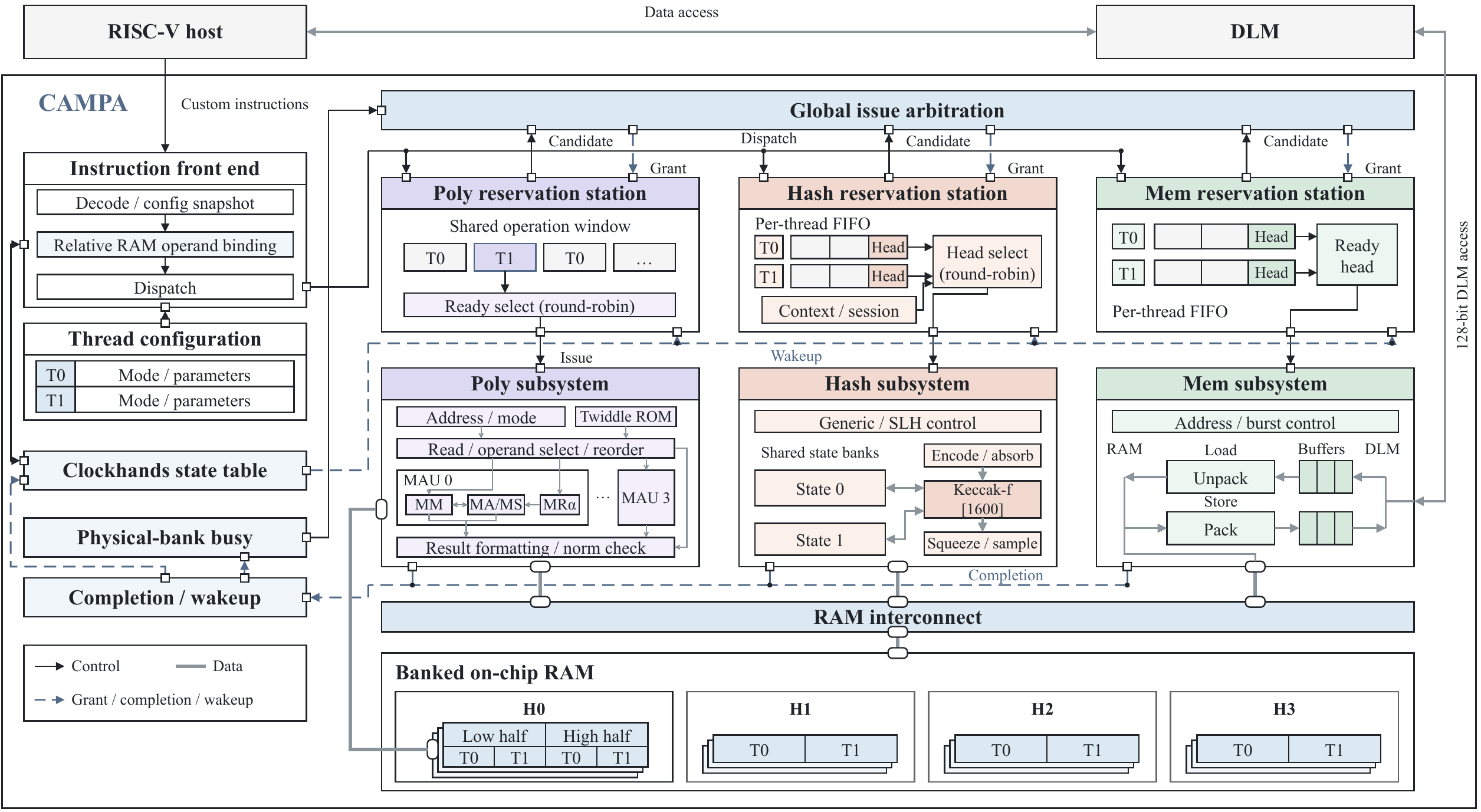}
  \caption{Overall architecture of \campa.}
  \label{fig:architecture-overview}
\end{figure*}

\subsection{Overall Architecture}\label{sec:overall-architecture}

\campa keeps independent tasks in hardware threads on shared Poly, Hash, Mem, and banked RAM resources (Fig.~\ref{fig:architecture-overview}). Threads retain task configuration, operand identities, and completion state; instructions carry thread identifiers. Dispatch binds operands and queues operations; ready work issues to available units, and completion updates versions and wakes consumers. Poly combines modular arithmetic units (MAUs), reordering, formatting, and norm checking. Hash combines encoding/absorption, one Keccak datapath, squeezing/sampling, and two state banks. Mem loads, stores, packs, and copies through a 128-bit data local memory (DLM) interface with exclusive host/accelerator ownership handoff.

Software configures threads and interleaves their instructions. Before a control decision or result access, a check-and-wait instruction returns the selected thread's status after its earlier operations complete and the queried RAM version is safe to reuse. The shared instruction input stalls while queued work from all threads continues executing. Software then advances, retries, or saves outputs and releases contexts before thread reuse.

\subsection{RAM Versioning and Dependency Tracking}\label{sec:ram-dependencies}

Interleaved tasks require stable operand identities and protection against premature RAM overwrite. RAM comprises four groups, or \emph{hands}, H0--H3, each with eight banks providing a 64-word, 96-bit RAM-block slot per thread. Threads share physical ports. H1--H3 banks are single-port; H0 regions split into low/high single-port halves accessed through the interconnect and a direct Poly path.

\begin{figure*}[t]
  \centering
  \begin{minipage}[t]{0.49\textwidth}
    \centering
    \includegraphics[width=\linewidth]{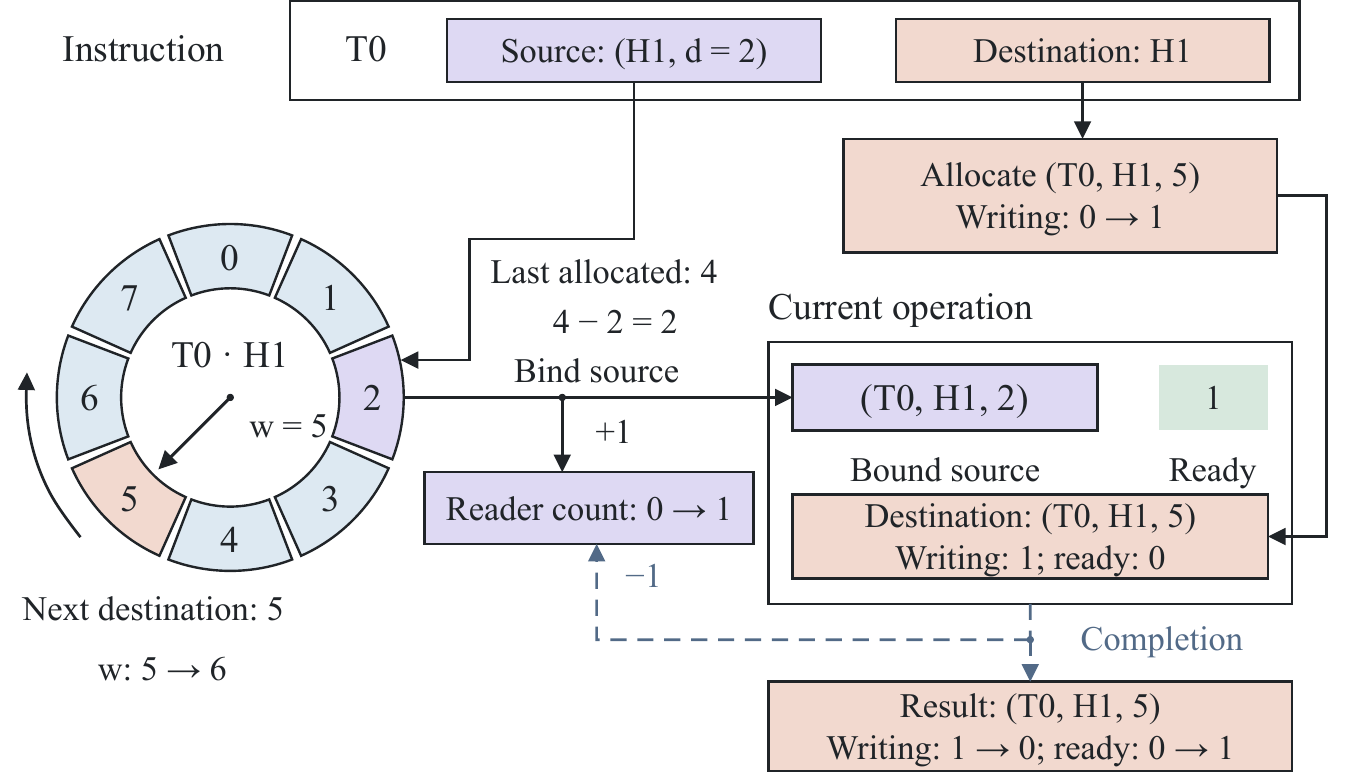}
    \par\smallskip
    {\footnotesize (a) RAM operand binding and version lifetime}
  \end{minipage}\hfill
  \begin{minipage}[t]{0.49\textwidth}
    \centering
    \includegraphics[width=\linewidth]{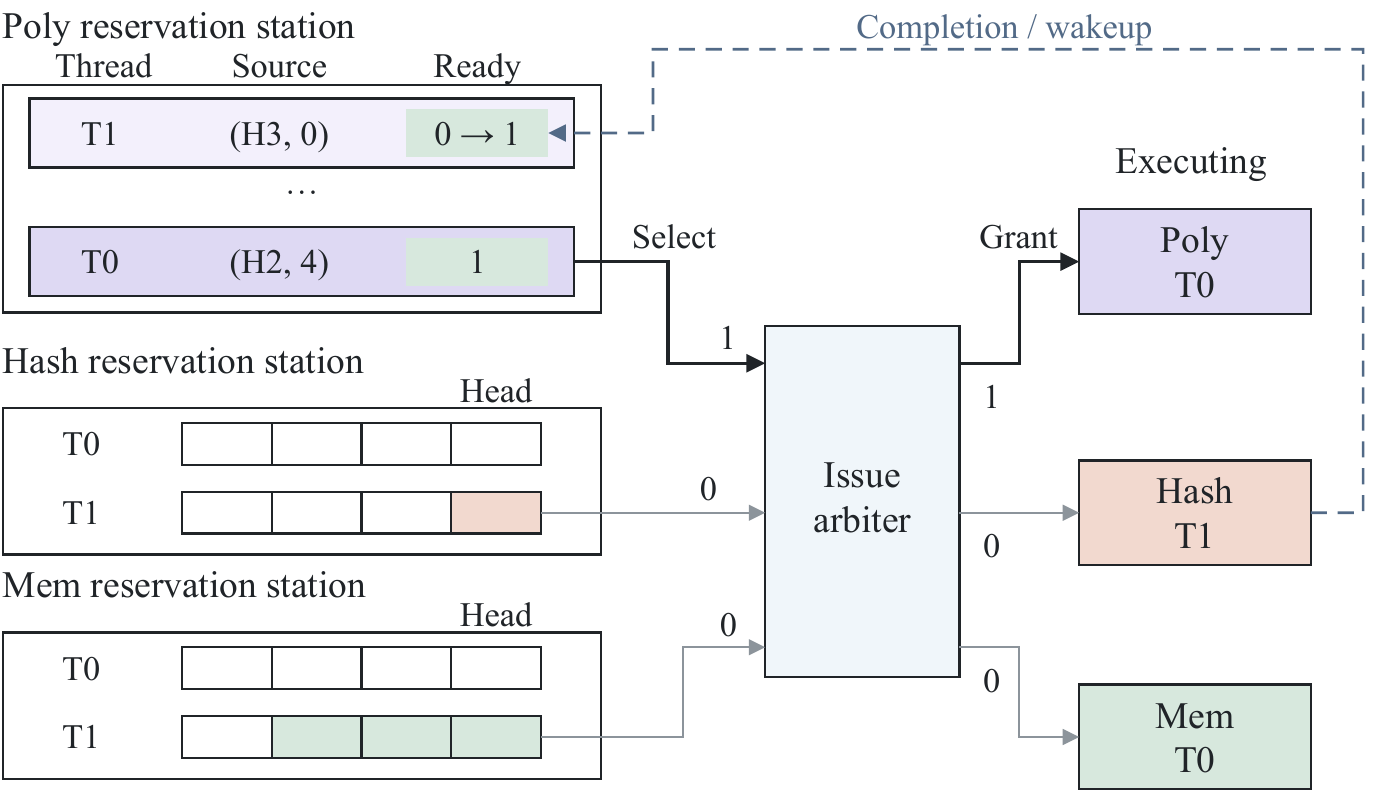}
    \par\smallskip
    {\footnotesize (b) Reservation-station scheduling}
  \end{minipage}
  \caption{RAM version binding and inter-task scheduling.}
  \label{fig:ram-binding-scheduling}
\end{figure*}

Adapting Clockhands~\cite{clockhands}, each thread $t$ maintains a write pointer $w_{t,h}$ per hand. A source distance $d$ resolves to slot $s$:
\begin{equation}
s=(w_{t,h}-1-d)\bmod 8.
\label{eq:clockhands-source}
\end{equation}
A destination takes the next slot and advances its hand's pointer at allocation. Sources bind $(t,h,s)$ identities before queueing, so later pointer advances and execution reordering preserve their operands (Fig.~\ref{fig:ram-binding-scheduling}a). Independent hands accommodate different update rates without a general mapping table or free-slot FIFO.

Allocation marks a destination writing/not-ready and registers its consumers as readers, including consumers queued before production. Producer completion marks it ready and wakes consumers; consumer completion releases readers. A wrapping writer waits for the prior writer and all readers, protecting data from binding through consumption. Separate physical-bank busy bits arbitrate ports across threads at issue. Software copies or stores values needed beyond the circular allocation window before reuse.

In-place NTT/INTT retains old and new epochs of the source slot while using the allocated destination as scratch. Allocation binds the transform to the old epoch and subsequent consumers to a pending new epoch. The transform waits for old-value readiness and earlier readers; completion retires the old epoch and wakes new-value consumers. Reuse requires both epochs free and no pending transform. Thus logical data lifetime and physical-port occupancy remain distinct constraints.

\begin{figure*}[t]
  \centering
  \begin{minipage}[b]{\columnwidth}
    \renewenvironment{figure}[1][]{}{}
\begin{figure}[t]
  \centering
  \includegraphics[width=\columnwidth]{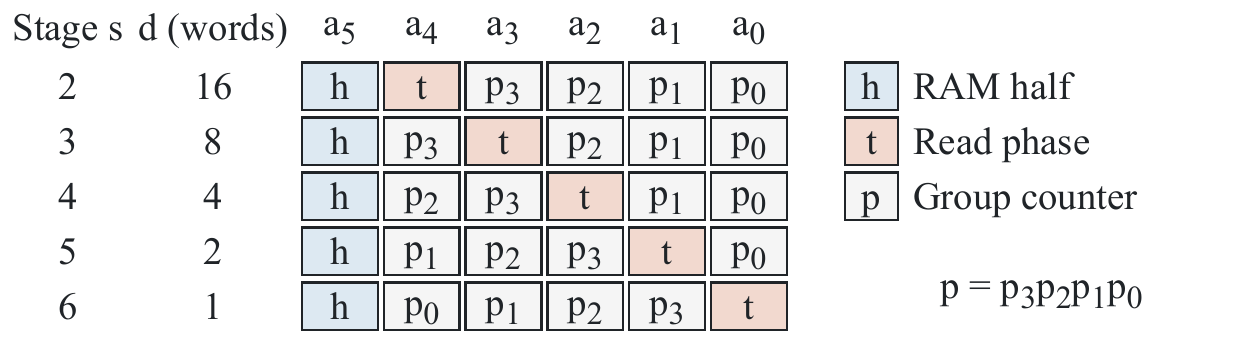}
  \caption{Stage-dependent address generation for \mldsa NTT.}
  \label{fig:ntt-execution}
\end{figure}
  \end{minipage}\hfill
  \begin{minipage}[b]{\columnwidth}
    \renewenvironment{figure}[1][]{}{}
\begin{figure}[t]
  \centering
  \includegraphics[width=\columnwidth]{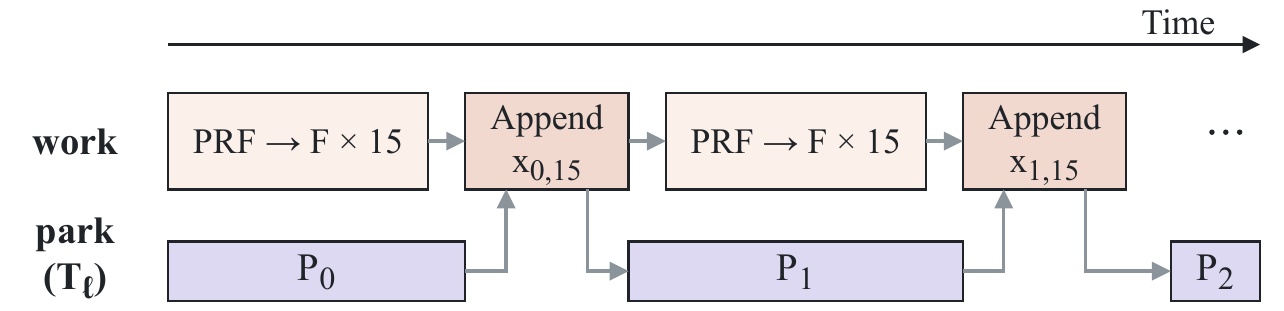}
  \caption{State retention across WOTS+ chains.}
  \label{fig:hash-state-execution}
\end{figure}
  \end{minipage}
\end{figure*}

\subsection{Cross-Algorithm Dynamic Scheduling}\label{sec:dynamic-scheduling}

Poly and Hash capture algorithm configurations at dispatch so ready operations from different algorithms share execution resources. Scheduling combines local candidate selection with global RAM arbitration (Fig.~\ref{fig:ram-binding-scheduling}b). Poly's shared reservation station~\cite{tomasulo} scans all threads' entries with a rotating pointer advanced on successful issue. Candidates need ready source versions, free banks, and an available executor; in-place transforms also await old-epoch readers. Blocked entries retain bindings while independent work proceeds.

Hash and Mem use per-thread FIFOs to preserve local sequencing. Hash selects an eligible head after the thread's preceding command completes and context/ownership admission succeeds. Mem selects among ready heads; both use rotating thread preference, with Hash additionally subject to ownership. A module requests only when its executor can accept work. The global arbiter rechecks banks and grants at most one operation per cycle with Hash--Poly--Mem priority. Issued operations continue concurrently; completion releases banks, updates versions, and wakes consumers.

In Fig.~\ref{fig:ram-binding-scheduling}b, T1's Poly operation waits for Hash data while T0's ready Poly receives the grant. Previously issued Hash and Mem operations keep running. Hash completion makes T1's source ready; T1 then issues when Poly is available. Cross-task readiness therefore fills execution windows without changing per-task dependencies.

\subsection{Role-Reconfigurable Keccak State Management}\label{sec:keccak-state}

\campa reuses two 1600-bit state banks across lattice hashing and \slhdsa with one Keccak datapath. Lattice mode retains independent resident/parked contexts with owner, function, byte offset, and absorption/finalization metadata. Commands can complete with contexts retained; selecting a parked context exchanges state and metadata to resume it. In \slhdsa mode, work executes short hashes while park retains a SHAKE stream or incremental-compression state.

Round groups read/write their owning bank directly; context tags preserve destinations through bank exchanges, avoiding another 1600-bit working copy. Within \slhdsa, begin/append retains thread ownership and termination releases it. Across modes, software hands over the bank pair at a quiescent boundary after retiring live contexts/session state; a new hash clears or constructs its state. Sec.~\ref{sec:operator-execution} describes operator execution using these mechanisms.
\section{Architecture-Aware Operator Execution}\label{sec:operator-execution}

\subsection{NTT Address Generation over Shared RAM}\label{sec:ntt-execution}

Stage-dependent accesses motivate bank-aware NTT mappings~\cite{scalable-conflict-free-ntt,ntt-conflict-graph}. A transform command executes Algorithm~\ref{alg:forward-ntt} by alternating source and working slots. Reusing both slots removes per-stage submission and destination allocation; the version interface publishes the final result. Each H0 slot contains 64 96-bit words in low/high single-port halves. In four-MAU \mldsa, a word holds four 24-bit coefficients. Shorter spans place paired words in one half, serializing reads. \campa reads both halves together and buffers word pairs.

Fig.~\ref{fig:ntt-execution} shows how forward stages 2--6 form word addresses by concatenating the half-select bit $h$, read-phase bit $t$, and four group-counter bits $p$. The phase bit moves to a lower address position as the butterfly span decreases. Each phase accesses both RAM halves; two phases collect a butterfly word pair from each half. Buffers feed these pairs to the MAUs and regroup writeback, retaining address/twiddle tags. Sixteen groups cover 64 words in 32 paired reads and 32 paired writes.

The first stage pairs halves directly; the last two form coefficient pairs within words. \mlkem selects its packed-word mapping and physical-half selector. These mappings reuse H0 ports and MAUs for both schemes without a separate polynomial memory.

\subsection{State-Retaining Hash Execution}\label{sec:hash-execution}

\begin{figure*}[t]
  \centering
  \resizebox{0.78\textwidth}{!}{%
  \begin{minipage}{\textwidth}
  \centering
\subfloat[\textbf{Throughput across request combinations.} Main configuration. Off-diagonal entries alternate 64 requests of each operation, starting with the row operation; diagonal entries contain 128 requests of a single operation. Throughput is normalized to serial execution of the same request sequence.\label{fig:evaluation-mixed}]{%
\begin{minipage}[b]{0.488\textwidth}
  \centering
  \includegraphics[trim=12bp 23bp 2bp 11bp,clip,width=0.85\linewidth]{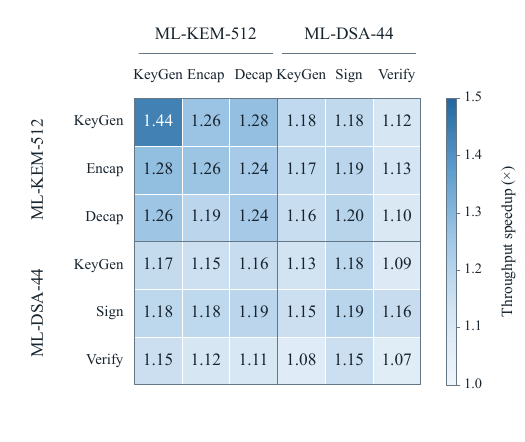}
\end{minipage}}\hfill
\subfloat[\textbf{Batch and thread scaling.} Throughput ratios relative to each configuration's T2 serial batch at the same $N$ are computed per workload and then geometrically averaged.\label{fig:evaluation-scaling}]{%
\begin{minipage}[b]{0.488\textwidth}
  \centering
  \includegraphics[trim=4bp 3bp 4bp 3bp,clip,width=0.91\linewidth]{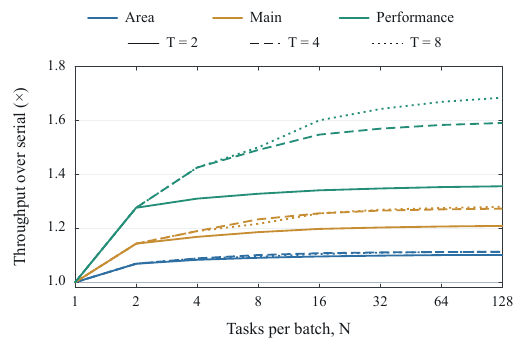}
\end{minipage}}

\smallskip

\subfloat[\textbf{Execution activity.} Arithmetic means across workloads.\label{fig:evaluation-activity}]{%
\begin{minipage}[b]{0.488\textwidth}
  \centering
  \includegraphics[width=\linewidth]{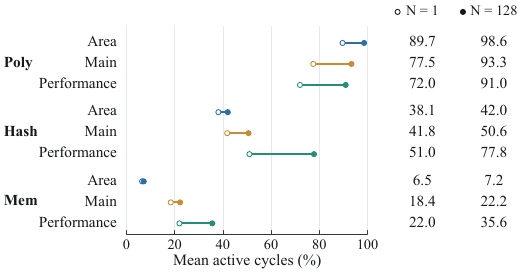}
\end{minipage}}\hfill
\subfloat[\textbf{Dynamic versus In-order selection.} Each workload's throughput is normalized to its configuration's In-order throughput at $N=1$; the resulting ratios are then geometrically averaged.\label{fig:evaluation-selection}]{%
\begin{minipage}[b]{0.488\textwidth}
  \centering
  \includegraphics[trim=4bp 3bp 4bp 3bp,clip,width=0.91\linewidth]{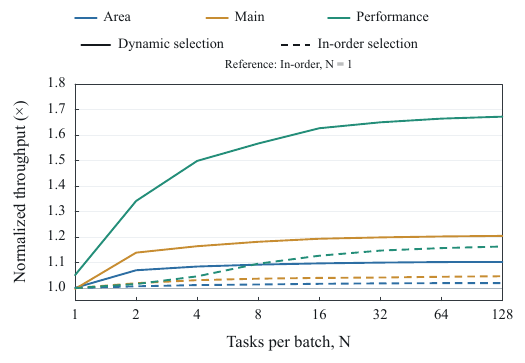}
\end{minipage}}
  \end{minipage}}
\caption{Throughput benefits of task concurrency and dynamic scheduling. Panels (b)--(d) cover 18 workloads: the three operations of each \mlkem-512/768/1024 and \mldsa-44/65/87 parameter set.}
\label{fig:evaluation-results}
\end{figure*}

For independent lattice-hash calls, the input path switches from A to B at A's permutation-wait boundary after outstanding transfers drain. B absorbs during A's permutation, and A resumes during B's. Saved function, offset, and progress preserve each stream; one Keccak core alternates permutations.

For \slhdsa, repeated short hashes make input preparation and node transfers part of the execution cost. Writing WOTS+ chain endpoints to RAM before compression introduces collection and reread traffic. \campa connects endpoint generation directly to incremental compression (Fig.~\ref{fig:hash-state-execution}). Seed registers retain recurring inputs, and an internal ADRS generator constructs node addresses, reducing repeated memory reads and software input preparation.

In \campa's WOTS+ public-key generation~\cite{sphincs-framework}, PRF initializes chain $i$; with $w=16$, $F$ runs 15 times to produce $x_{i,15}$~\cite{nist-fips-205}. Work executes PRF/$F$, while park retains the $T_\ell$ accumulator $P_i$ after its prefix and first $i$ endpoints:
\begin{equation}
P_{i+1}=\operatorname{Absorb}(P_i,x_{i,15}),\quad
\mathrm{pk}_{\mathrm{WOTS}}=\operatorname{Squeeze}(P_\ell).
\label{eq:wots-retained-context}
\end{equation}
Endpoint absorption restores $P_i$ to work and commits $P_{i+1}$ to park before the next chain. The internal endpoint path eliminates the extra RAM round trip used to gather compression inputs. The same retained-state organization supports FORS-root compression, protected by Sec.~\ref{sec:keccak-state}'s ownership protocol.
\section{Evaluation}\label{sec:evaluation}

\subsection{Experimental Setup}

To examine how compute and thread resources affect throughput and utilization, we evaluate three CAMPA configurations: Area, Main, and Performance. Area and Main each use one Poly executor, with one and four MAUs, respectively. Area's Hash unit executes one Keccak round per cycle and uses one sampler lane; Main executes four rounds per cycle with four sampler lanes. Performance doubles Main's Poly executors while retaining its Hash resources. Unless otherwise specified, Area and Main support up to two resident tasks ($T=2$), and Performance supports four ($T=4$), with one task per hardware thread.

\begin{table*}[t]
  \centering
  \begin{minipage}[t]{\columnwidth}
    \vspace{0pt}
    \renewenvironment{table}[1][]{}{}
\begin{table}[t]
  \caption{FPGA resource comparison of PQC accelerators.}
  \label{tab:prior-fpga}
  \centering
  \footnotesize
  \setlength{\tabcolsep}{3pt}
  \begin{tabular*}{\columnwidth}{@{\extracolsep{\fill}}llrrrrr@{}}
    \toprule
    Work & Platform & LUT & FF & DSP & \shortstack{BRAM} & kENS \\
    \midrule
    \cite{kali} & Zynq US+ & 23,347 & 9,798 & 4 & 24  & 12.262 \\
    \cite{horcrux-complete} & Zynq US+ & 20,196 & 4,429 & 11 & 0  & 6.703 \\
    \cite{lighthd} & Artix-7 & 20,623 & 13,854 & 8 & 12  & 10.088 \\
    \cite{sloth} & Artix-7 & 14,428 & NR & 0 & 32  & NR \\
    \midrule
    \textbf{CAMPA main} & Zynq US+ & 89,275 & 17,989 & 8 & 0 & 25.367 \\
    \textbf{CAMPA area} & Zynq US+ & 65,859 & 14,581 & 2 & 0 & 18.487 \\
    \textbf{CAMPA perf.} & Zynq US+ & 127,074 & 25,744 & 16 & 0 & 36.587 \\
    \bottomrule
  \end{tabular*}
  \par\smallskip
  \begin{minipage}{\columnwidth}
    \footnotesize
    ENS~\cite{lighthd}: $\mathrm{LUT}/4+\mathrm{FF}/8+100\,\mathrm{DSP}+200\,\mathrm{BRAM}$.
    \par\smallskip\noindent
    CAMPA LUT breakdown (logic/storage): Main, 75,981/13,294; Area, 52,495/13,364; Performance, 101,160/25,914.
  \end{minipage}
\end{table}
  \end{minipage}\hfill
  \begin{minipage}[t]{\columnwidth}
    \vspace{0pt}
    \renewenvironment{table}[1][]{}{}
\begin{table}[t]
  \caption{ASIC resource comparison of PQC accelerators.}
  \label{tab:prior-asic}
  \centering
  \footnotesize
  \setlength{\tabcolsep}{3pt}
  \begin{tabular*}{\columnwidth}{@{\extracolsep{\fill}}llrr@{}}
    \toprule
    Work & Process & Area (mm$^2$) & Logic (kGE) \\
    \midrule
    \cite{matrix-riscv} & 28 nm & NR & 581.4 \\
    \cite{unified-pqc-soc} & 28 nm & 0.2574 & NR \\
    \cite{horcrux-complete} & 65 nm & NR & 115.8 \\
    \cite{kem22} & 22 nm & 0.0629$^{\rm M}$ & NR \\
    \midrule
    \textbf{CAMPA main} & 28 nm & 0.1876 & 496.3 \\
    \textbf{CAMPA area} & 28 nm & 0.1802 & 476.7 \\
    \textbf{CAMPA perf.} & 28 nm & 0.2839 & 751.1 \\
    \bottomrule
  \end{tabular*}
  \par\smallskip
  \begin{minipage}{\columnwidth}
    \footnotesize
$^{\rm M}$includes SRAM.
  \end{minipage}
\end{table}
  \end{minipage}

  \medskip

  \begingroup
    \renewenvironment{table*}[1][]{}{}
    \begin{table*}[t]
\caption{Throughput-equivalent time and full-flow ATP.}
\label{tab:prior-performance}
\centering\footnotesize
\setlength{\tabcolsep}{2pt}
\begin{tabular*}{\textwidth}{@{\extracolsep{\fill}}lllrrrrr@{}}
\toprule
Algorithm & Work & Platform & Freq. (MHz) & KeyGen ($\mu$s) & Enc./Sign ($\mu$s) & Dec./Verify ($\mu$s) & Flow ATP \\
\midrule
\multirow{7}{*}{ML-KEM-512} & \cite{horcrux-complete} & Zynq US+ & 42 & 3,880.95 & 4,500 & 6,904.76 & 102,000 \\
 & \cite{kali} & Zynq US+ & 270 & 12.57 & 18.36 & 25.21 & 688 \\
 & \textbf{CAMPA} & Zynq US+ & 105.5/128.7 & 12.80/\textbf{8.76} & 20.38/\textbf{9.86} & 28.64/\textbf{12.90} & 1,570/1,150 \\
\cmidrule(l){2-8}
 & \cite{matrix-riscv} & 28 nm & 540 & 17.41 & 35.19 & 81.11 & (77.7; \textemdash) \\
 & \cite{unified-pqc-soc} & 28 nm & 700 & \multicolumn{3}{c}{11.77 (full flow)} & (\textemdash; 3.03) \\
 & \cite{kem22} & 22 nm & 938 & 5.58 & 7.41 & 10.09 & \textemdash \\
 & \textbf{CAMPA} & 28 nm & 800 & 1.69/\textbf{1.41} & 2.69/\textbf{1.59} & 3.78/\textbf{2.07} & (4.04; 1.53)/(\textbf{3.81}; \textbf{1.44}) \\
\midrule
\multirow{7}{*}{ML-DSA-44} & \cite{horcrux-complete} & Zynq US+ & 42 & 10,904.76 & 36,595.24 & 14,095.24 & 413,000 \\
 & \cite{kali} & Zynq US+ & 270 & 54.05 & 117.27 & 57.12 & 2,800 \\
 & \cite{lighthd} & Artix-7 & 212.8 & 19.7 & 57.3 & 25.8 & 1,040 \\
 & \textbf{CAMPA} & Zynq US+ & 105.5/128.7 & 55.28/30.63 & 167.36/83.41 & 75.60/33.62 & 7,570/5,400 \\
\cmidrule(l){2-8}
 & \cite{matrix-riscv} & 28 nm & 540 & 84.81 & 324.26 & 166.3 & (335; \textemdash) \\
 & \cite{unified-pqc-soc} & 28 nm & 700 & \multicolumn{3}{c}{42.87 (full flow)} & (\textemdash; 11.0) \\
 & \textbf{CAMPA} & 28 nm & 800 & 7.29/\textbf{4.93} & 22.06/\textbf{13.42} & 9.97/\textbf{5.41} & (19.5; 7.38)/(\textbf{17.8}; \textbf{6.74}) \\
\midrule
\multirow{4}{*}{\shortstack[l]{SLH-DSA-\\SHAKE-128f}} & \cite{sloth} & Artix-7 & 100 & 1,765.52 & 49,039.78 & 4,406.36 & \textemdash \\
 & \textbf{CAMPA} & Zynq US+ & 105.5/128.7 & 333.09/\textbf{272.92} & 21,866.36/\textbf{17,916.44} & 901.43/\textbf{738.59} & 586,000/693,000 \\
\cmidrule(l){2-8}
 & \cite{unified-pqc-soc} & 28 nm & 500 & \multicolumn{3}{c}{779.42 (full flow)} & (\textemdash; 201) \\
 & \textbf{CAMPA} & 28 nm & 800 & 43.91 & 2,882.31 & 118.82 & (1,510; 571)/(2,290; 865) \\
\bottomrule
\end{tabular*}
\par\smallskip
\begin{minipage}{\textwidth}\footnotesize
CAMPA pairs: Main/Performance; a single value applies to both. CAMPA lattice times: $N=128$ batch means; other per-operation times: serial references.
ATP (three significant digits): FPGA kENS$\cdot\mu$s; ASIC (MGE$\cdot\mu$s; mm$^2\cdot\mu$s).
\end{minipage}
\end{table*}
  \endgroup
\end{table*}

We synthesize the RTL using Synopsys Design Compiler in SMIC 28-nm technology at the TT corner, targeting 800~MHz. FPGA implementation uses Vivado 2024.2 on the Zynq UltraScale+ XCZU7EV in an out-of-context flow, with frequencies obtained from post-route timing analysis.

\subsection{Task Concurrency and Dynamic Scheduling}\label{sec:mixed-results}\label{sec:thread-results}\label{sec:activity-results}\label{sec:selection-results}

Concurrent requests achieve $1.07$--$1.44\times$ throughput over serial execution across the combinations in Fig.~\ref{fig:evaluation-results}(a), including mixtures of operations and algorithms. A broader evaluation of 81 \mlkem/\mldsa pairings yields geometric-mean gains of $1.17\times$/$1.35\times$ for Main/Performance ($T=2$), with all pairs benefiting.

Richer compute configurations benefit more from additional hardware threads and the resulting pool of ready work (Fig.~\ref{fig:evaluation-results}(b)). At $N=128$, Performance's gains reach $1.36\times$/$1.59\times$/$1.68\times$ at $T=2/4/8$, while Area saturates near $1.11\times$.

Concurrency fills execution gaps (Fig.~\ref{fig:evaluation-results}(c)). At $N=128$, Area's 98.6\% Poly activity leaves little idle capacity, consistent with its thread-scaling saturation. Performance's larger Poly and thread budgets enable greater overlap: from $N=1$ to $N=128$, Hash activity rises from 51.0\% to 77.8\%, and Mem from 22.0\% to 35.6\%.

With matched hardware, instruction streams, and thread capacity, Dynamic can issue later ready candidates while In-order waits for the oldest admitted one (Fig.~\ref{fig:evaluation-results}(d)). This out-of-order selection improves $N=128$ throughput by 8.11\%, 15.12\%, and 43.78\% for Area/Main/Performance.

\subsection{Hardware Cost and Efficiency}\label{sec:hardware-cost}\label{sec:prior-comparison}

CAMPA uses LUTRAM to match its small, independently accessed RAM banks and avoid underfilled BRAM blocks. Table~\ref{tab:prior-fpga} includes logic and storage LUTs; ASIC ATP uses CAMPA's engine-logic area (Tables~\ref{tab:prior-asic} and~\ref{tab:prior-performance}).

At 28~nm, Main achieves $1.98\times$/$1.50\times$ throughput per engine-logic area over VLSI'25~\cite{unified-pqc-soc} for ML-KEM-512/ML-DSA-44; Performance reaches $2.10\times$/$1.64\times$. On Zynq UltraScale+, Performance reduces these schemes' summed throughput-equivalent times by 43.9\% and 35.4\%, respectively, relative to KaLi~\cite{kali}. ML-DSA Sign is best-case for CAMPA and~\cite{unified-pqc-soc,kali,lighthd}.

Shared Keccak and retained states support complete SLH-DSA-SHAKE-128f operations. Matching Hash resources and ASIC clocks give Main and Performance identical SLH times; Performance's additional Poly resources target \mlkem/\mldsa throughput.
\section{Conclusion}
\label{sec:conclusion}

\campa brings cross-algorithm multithreading and inter-task dynamic scheduling
to a shared \riscv PQC coprocessor supporting \mlkem, \mldsa, and \slhdsa.
Across 18 lattice-PQC workloads, architectural simulation shows geometric-mean
throughput gains of $1.21\times$ for Main T2 and $1.59\times$ for Performance
T4 over serial execution on the same hardware. By scheduling ready work across tasks, \campa turns multi-scheme
operator reuse into sustained utilization of shared computation and storage.
\section*{Content Generated by AI}

GPT-6.1 Sol assisted with manuscript drafting, language polishing and translation.
The authors reviewed generated content and take
responsibility for all claims and results.

\clearpage
\IEEEtriggeratref{19}
\bibliographystyle{IEEEtran}
\bibliography{references}

\end{document}